\documentclass[
reprint,
superscriptaddress,
aps,
prl,
]{revtex4-2}

\usepackage{graphicx}
\usepackage{dcolumn}
\usepackage{bm}
\usepackage[utf8]{inputenc}
\usepackage{amsmath, amsthm, amssymb, amsfonts}
\usepackage{accents}
\usepackage[usenames,dvipsnames]{xcolor}
\usepackage[caption=false]{subfig}
\usepackage{tikz}
\usepackage{array}
\usepackage{dsfont}
\usepackage{mathtools}
\usepackage{soul,xcolor}
\usepackage[colorlinks=true,linkcolor=Blue,citecolor=Blue,urlcolor=Blue]{hyperref}
\usepackage[capitalise]{cleveref}

\usepackage[version=4]{mhchem}
\usepackage{siunitx}

\DeclarePairedDelimiter\floor{\lfloor}{\rfloor}

\newcommand{\YZ}{\color{black}}

\setstcolor{magenta}

\begin{document}

\title{
Basins with Tentacles
}

\author{Yuanzhao Zhang}
\affiliation{Center for Applied Mathematics, Cornell University, Ithaca, New York 14853, USA}
\affiliation{Santa Fe Institute, 1399 Hyde Park Road, Santa Fe, New Mexico 87501, USA}
\author{Steven H. Strogatz}
\affiliation{Center for Applied Mathematics, Cornell University, Ithaca, New York 14853, USA}
\affiliation{Department of Mathematics, Cornell University, Ithaca, New York 14853, USA}

\begin{abstract}
To explore basin geometry in high-dimensional dynamical systems, we consider a ring of identical Kuramoto oscillators. Many attractors coexist in this system; each is a twisted periodic orbit characterized by a winding number $q$, with basin size proportional to $e^{-kq^2}.$ We uncover the geometry behind this size distribution and find the basins are octopus-like, with nearly all their volume in the tentacles, not the head of the octopus (the ball-like region close to the attractor). We present a simple geometrical reason why basins with tentacles should be common in high-dimensional systems.

\vspace{3mm}
\noindent DOI: \href{https://doi.org/10.1103/PhysRevLett.127.194101}{10.1103/PhysRevLett.127.194101} 
\end{abstract}


\maketitle


Basins of attraction are fundamental to the analysis of dynamical systems \cite{milnor1985concept,aguirre2009fractal,ott2002chaos}.
Over the years, many remarkable properties of basins have been discovered \cite{crutchfield1988subbasins,crutchfield1988attractors,wiesenfeld1989attractor,kaneko1990clustering,kaneko1997dominance,timme2002prevalence}, most notably that their geometry can be wild, as exemplified by Wada basins \cite{nusse1996basins}, fractal basin boundaries \cite{mcdonald1985fractal,grebogi1987chaos}, and riddled or intermingled basins \cite{alexander1992riddled,sommerer1993physical,ott1993scaling,heagy1994experimental,ashwin1996attractor}.
Yet despite these foundational studies, much remains to be learned about basins,
especially in systems with many degrees of freedom.
Even the simplest questions---how big are the basins, and what shapes do they have?---constitute an active area of research \cite{wiley2006size,xu2011direct,monteforte2012dynamic,menck2013basin,menck2014dead,zou2014basin,martiniani2016turning,martens2016basins,leng2016basin,mitra2017multiple,delabays2017size,schultz2017potentials,rakshit2017basin,martiniani2017numerical,belykh2019synchronization,zhang2020critical,townsend2020dense,andrzejak2020two,andrzejak2021chimeras}.

Given the rich properties of basins, it is perhaps not surprising that even for locally-coupled Kuramoto oscillators---arguably one of the simplest dynamical systems on networks---there are still riddles to be solved.
In one study, Wiley, Strogatz, and Girvan~\cite{wiley2006size} numerically investigated a ring of $n$ identical Kuramoto oscillators and measured the basin sizes of its coexisting attractors. All the attractors in this system could be conveniently characterized by an integer $q$, because they were all  phase-locked periodic orbits in which the oscillators' phases formed $q$ full twists around the ring. By sampling the entire state space uniformly, these authors found that the basin sizes for the $q$-twisted states were Gaussian distributed as a function of the winding number $q$. 

More recently, Delabays, Tyloo, and  Jacquod~\cite{delabays2017size} measured the basin sizes of the same system in a different way.
For each $q$, they estimated
the distance from the $q$-twisted state to its basin boundary. Then they used that distance to calculate the volume of an associated hypercube, which was taken as a proxy for the basin size.  
Intriguingly, they reached a completely different conclusion and reported that the basin sizes decreased exponentially with $|q|$.
The tension between the two results created a puzzle that underscored our lack of understanding of basins in even the simplest systems.

In this Letter, we resolve the tension by showing that high-dimensional basins tend to have convoluted geometries 
and cannot be approximated by simple 
shapes such as hypercubes.
Although they are impossible to visualize precisely (because of their high dimensionality), we present evidence that these basins have long tentacles that reach far and wide and become tangled with each other. 
Yet sufficiently close to its own attractor, each basin becomes rounder and more simply structured, somewhat like the head of an octopus \footnote{The ``dendritic” structure of the basins for \cref{eq:kuramoto_nn} was first noted by Stefano Martiniani in his Ph.D. thesis (\url{https://doi.org/10.17863/CAM.12772}).}.

Returning to the issue of the basin sizes, we find that their volumes are proportional to $e^{-kq^2}$, not $e^{-{k|q|}}$. 
The  discrepancy can be traced to how the latter distribution was obtained~\cite{delabays2017size}: it was based on local measurements and thereby ignored the basin's tentacles. 
Such estimates miss nearly all of a basin's volume, even for moderate network sizes. 
In terms of our metaphor, almost all of a basin's volume is in its tentacles, not its head. 

This finding is not limited to Kuramoto oscillators. 
We provide a simple geometrical argument showing that, as long as the number of attractors in a system grows sub-exponentially with system size, the basins are expected to be octopus-like. 
As further evidence of their genericity, basins of this type were previously found in simulations of jammed sphere packings \cite{ashwin2012calculations,martiniani2016structural}, where they were described as ``branched'' and ``threadlike'' away from a central core \cite{ashwin2012calculations} and accurate methods were developed for computing their volumes \cite{xu2011direct,asenjo2014numerical, martiniani2016structural}.
There is also enticing evidence of octopus-like basins in neuronal networks \cite{monteforte2012dynamic}, power grids \cite{menck2014dead}, and photonic couplers \cite{zhiyenbayev2019enhanced}.


The equations for our Kuramoto model on a ring are 
\begin{equation}
	\dot{\theta}_i = \sin(\theta_{i+1}-\theta_i) + \sin(\theta_{i-1}-\theta_i)
	\label{eq:kuramoto_nn}
\end{equation}
for $i = 1, \ldots, n$. Here $\theta_i(t) \in S^1$ is the phase of oscillator $i$ at time $t$ and $n$ is the number of oscillators. For convenience, we have written the equations in a frame rotating at the common frequency $\omega$ of all the oscillators; without loss of generality, we have set $\omega=0$ by going into this frame. We assume a periodic boundary condition,  $\theta_{n+1}(t) = \theta_1(t)$ mod $2 \pi$, since the oscillators are assumed to be connected in a ring. 
One can show that the in-phase synchronous state $\theta_i(t)=\theta_j(t)\;\forall\, i,j,t$ is always an attractor for \cref{eq:kuramoto_nn}.
But if $n$ is large, there are also many other competing attractors \cite{dorfler2013synchronization,delabays2016multistability,manik2017cycle}. In all of these, the oscillators are phase locked and run at the same instantaneous frequency, $\dot{\theta}_i=\dot{\theta}_j\;\forall\, i,j$ (which can be set to zero in the rotating frame we are using). In these states, the phases of the oscillators make $q$ full twists around the ring and satisfy $\theta_i = 2 \pi i q/n + C,$  where $q$ is the winding number of the state~\cite{wiley2006size}. These twisted states exist for all $q$, but only those with $|q|<n/4$ are stable~\cite{delabays2016multistability,manik2017cycle,delabays2017size,townsend2020dense}.

A natural question is then: What are the basin sizes for each of the stable twisted states \cite{wiley2006size,ha2012basin,delabays2017size}? As mentioned above, this question has been explored from two perspectives, one global and one local. Before we present  our new results, we need to delve more deeply into both  perspectives, because understanding them will prove crucial to interpreting the numerical results presented below.

In Ref.~\cite{wiley2006size}, Wiley, Strogatz, and Girvan used a simple strategy of sampling initial conditions uniformly at random in the entire state space to estimate the basin sizes, and found that their volumes were  proportional to $e^{-kq^2}$.
However, even for a moderate number of oscillators, it would take an astronomical number of samples to cover the $n$-dimensional state space at a reasonable resolution (e.g., for $n=80$ used in Ref.~\cite{wiley2006size}, around $10^{80}$ points would be needed for a resolution of $10$ points in each direction, which is obviously completely infeasible). So one could well doubt that the results obtained by this procedure are meaningful, let alone reliable.

For this reason, Delabays, Tyloo, and  Jacquod~\cite{delabays2017size} designed a more sophisticated procedure to measure the basin sizes.
First they analytically calculated the distance from each twisted state to the nearest saddle point on its basin boundary~\cite{deville2012transitions} and found that distance to be proportional to $1-4|q|/n$.
Then, assuming that the basin is well approximated by an $n$-dimensional hypercube, the authors estimated the basin size to be proportional to $(1-4|q|/n)^n$, which approaches $e^{-4|q|}$ as $n\to \infty$.
This strategy enabled the authors to measure the basin sizes for twisted states with large winding numbers, which are extremely difficult to reach from uniform global sampling.

The two results, $e^{-kq^2}$ and $e^{-4|q|}$, cannot both be right. In fact, it seems quite possible that both could be wrong, since the number of samples used in Ref.~\cite{wiley2006size} could easily be insufficient to capture the correct scaling relation and the hypercube assumed in Ref.~\cite{delabays2017size} could easily be a poor approximation of the correct basin geometry.

\begin{figure}[t]
\centering
\includegraphics[width=1\columnwidth]{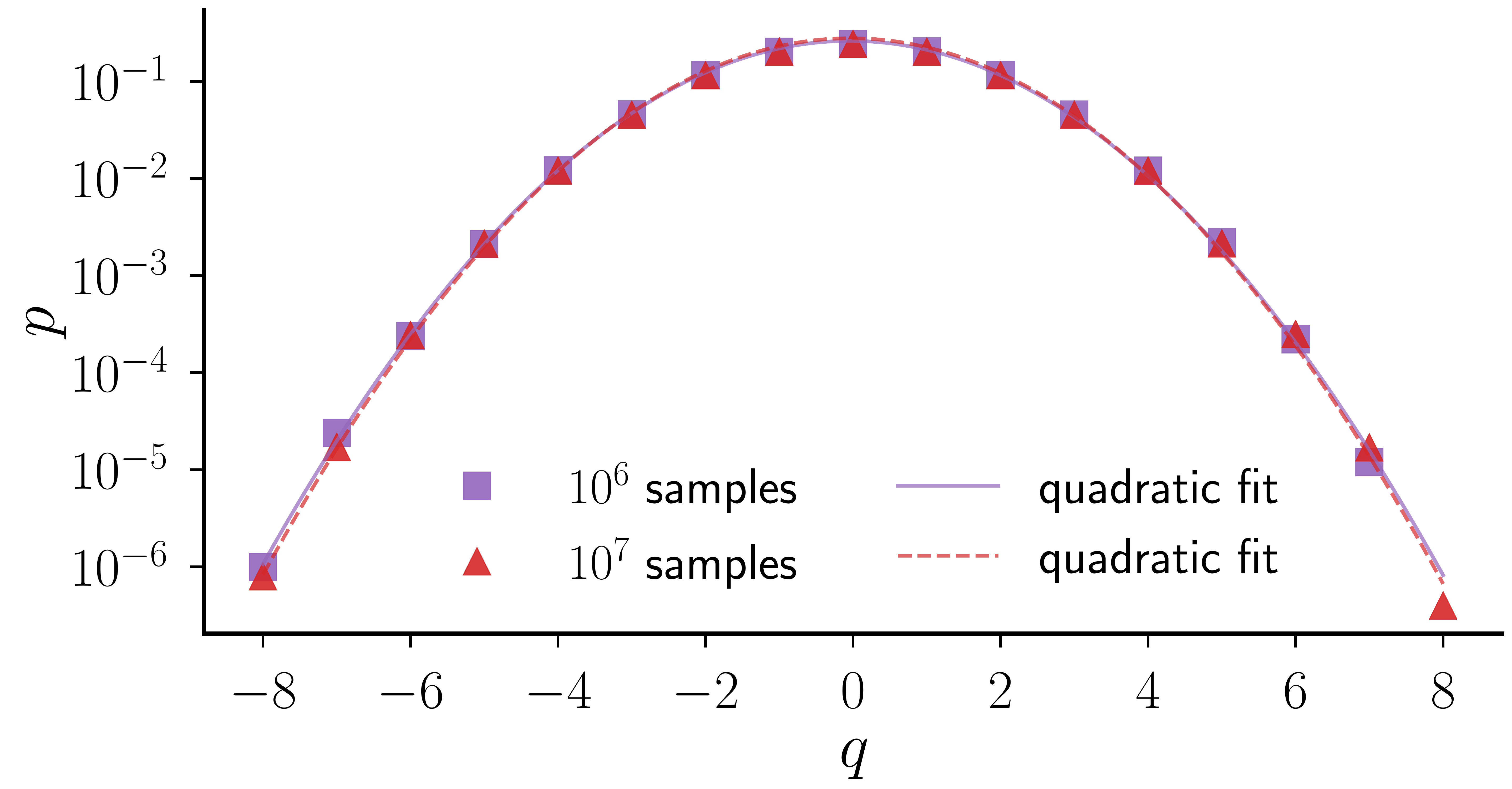}
\vspace{-6mm}
\caption{Probability of converging to the $q$-twisted states from random initial conditions. The system \eqref{eq:kuramoto_nn} consists of 83 identical Kuramoto oscillators with nearest-neighbor coupling.
}
\label{fig:1}
\end{figure}

To clarify what is going on here, we begin by testing convergence of the basin size estimates when we do finer and finer uniform global sampling.
Figure~\ref{fig:1} plots the relative basin sizes of the  $q$-twisted states for $n=83$ oscillators, as estimated by the measured probability $p$ of the states being reached from random initial conditions. 
At $N=10^7$ samples, the stable twisted states with $8<|q|\leq 20$ are completely undetected due to their minuscule basin sizes.
However, the estimates of basin sizes for twisted states with $|q|<7$ are clearly converged already for $N=10^6$ samples.

These numerical convergence results can be understood theoretically as follows. For each $q$-twisted state, a random point in the state space is either inside or outside of its basin.
Thus, using results from repeated Bernoulli experiments, the standard error of the relative basin size $p$ after $N$ samples is $\sqrt{p(1-p)}/\sqrt{N}$ \cite{menck2013basin,schultz2017potentials}.
For $p \ll 1$, the relative standard error is $1/\sqrt{pN}$, which is about $5\%$ for $p=10^{-3.5}$ (a value of $p$ that corresponds to $|q|=6$ in \cref{fig:1}) and $10^6$ samples.
Note that the relative standard error does not depend on the dimension $n$ of the state space, despite the increased difficulty of adequately covering the state space for larger $n$.
After confirming the convergence of the basin size estimates, we can see that the data strongly favor Gaussian dependence over exponential scaling, which is further supported by a least squares fit using quadratic functions.

\begin{figure}[t]
\centering
\includegraphics[width=1\columnwidth]{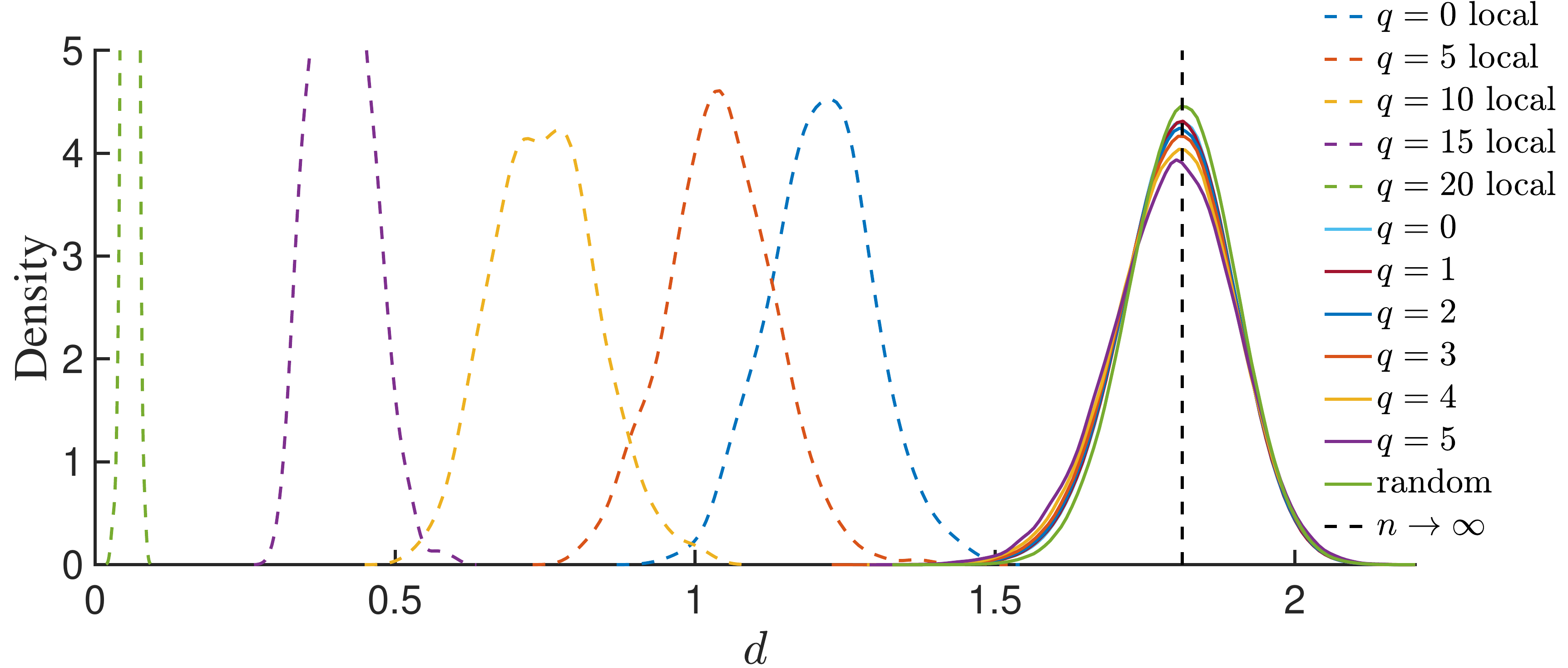}
\vspace{-6mm}
\caption{Local measurements of basins miss almost all of state space. The dashed curves represent the distributions of the distance away from a $q$-twisted state at which one first escapes its basin. The solid curves are the distributions of the distance between a $q$-twisted state and points inside its basin found through uniform global sampling. The solid curves for different $q$ collapse onto a single distribution, which matches the distribution of the distance between two randomly selected points in the state space. This distribution approaches a delta function at $d^{(\infty)}  \approx 1.81$ as $n\to \infty$ (black dashed line). If we use hypercubes to approximate basins based on dashed distributions (the strategy adopted in Ref.~\cite{delabays2017size}), then we omit almost all the points in all the basins. Simple calculations show that, for $n=83$ considered here, these hypercubes account for no more than $10^{-34}$ of the entire state space.
}
\label{fig:2}
\end{figure}

Next, we examine the hypercube assumption and develop an intuitive picture for the basin geometry.
Following Ref.~\cite{delabays2017size}, for many randomly selected directions we numerically identify the shortest distance away from the $q$-twisted state that is needed to escape the basin.
Specifically, for each $q$-twisted state, we incrementally increase the perturbation along a random direction until we find a point that is outside of the basin.
This procedure is repeated for $1000$ independent directions, and the dimension-normalized Euclidean distance $d = \left({\sum_{i=1}^n d_i^2/n}\right)^{1/2}$ between the escaping point and the attractor is recorded for each direction.
(Here, $d_i \in [0,\pi]$ is the absolute phase difference between the $i$th oscillators of the two states.)
The distributions of the escape distances are shown as dashed curves in \cref{fig:2}.
Indeed, similar to the results reported in Ref.~\cite{delabays2017size}, the distributions are all fairly concentrated with tight supports.
This suggests that the basins are fairly isotropic, which motivated the use of the hypercube approximation in Ref.~\cite{delabays2017size}.

However, when we compare these distributions with the data obtained from uniform global sampling, an inconsistency emerges.
For each sampled initial condition in \cref{fig:1}, we recorded the distance $d$ between the starting point and the attractor it converged to.
The solid curves in \cref{fig:2} show the distributions of these distances for $q$-twisted states with more than $10^4$ samples (data for states with $q<0$ are not shown as they mirror their $q>0$ counterparts).
One immediately notices that these distributions have much larger means than the dashed distributions and the two groups of distributions have essentially no overlap.
In other words, the majority of a basin is outside of the basin boundary identified through local perturbations!
This observation calls into question the hypercube assumption.
Indeed, if we take the means $\mu_q$ of the dashed distributions as the half side lengths of the hypercubes that approximate basins, the ratio between the volume of all hypercubes and the volume of the full state space is
\begin{equation}
	\frac{\sum_{q=-20}^{20}(2\mu_q)^{83}}{(2\pi)^{83}} \approx 10^{-34}.
\end{equation}
Note that $10^{-34}$ is still an overestimation of the ratio because of the inequality of arithmetic and geometric means.
Thus, the hypercubes based on the basin boundaries identified through local perturbations miss almost the entire state space.
This suggests that each basin must be ``leaking'' outside of its hypercube-like core and forms tentacles like an octopus.
Moreover, for large $n$,
most of the basin volume is concentrated in these tentacles.

Now, circling back to the distributions extracted from the globally sampled data (solid curves in \cref{fig:2}), we notice some additional interesting features.
First, the distributions for different $q$ almost completely overlap and seem to 
follow a master distribution.
Second, this master distribution agrees with the distribution of the distance between two randomly selected points in the state space (green solid curve).
This agreement implies that, statistically speaking, the points in a basin are distributed in the state space as if they were randomly placed and the basin profiles are in a sense uniform across different winding numbers $q$.

Moreover, as $n\to\infty$, the master distribution approaches a delta function at 
$d^{(\infty)} \approx 1.81$ (due to the central limit theorem).
In this limit, almost all points in the state space are $d^{(\infty)}$ away from any $q$-twisted state, and how the basins look outside the sphere with radius $d^{(\infty)}$ is irrelevant for determining basin sizes.
These considerations are not limited to Kuramoto oscillators and apply to fixed-point attractors in any high-dimensional dynamical system with a compact state space.

\begin{figure}[t]
\centering
\includegraphics[width=1\columnwidth]{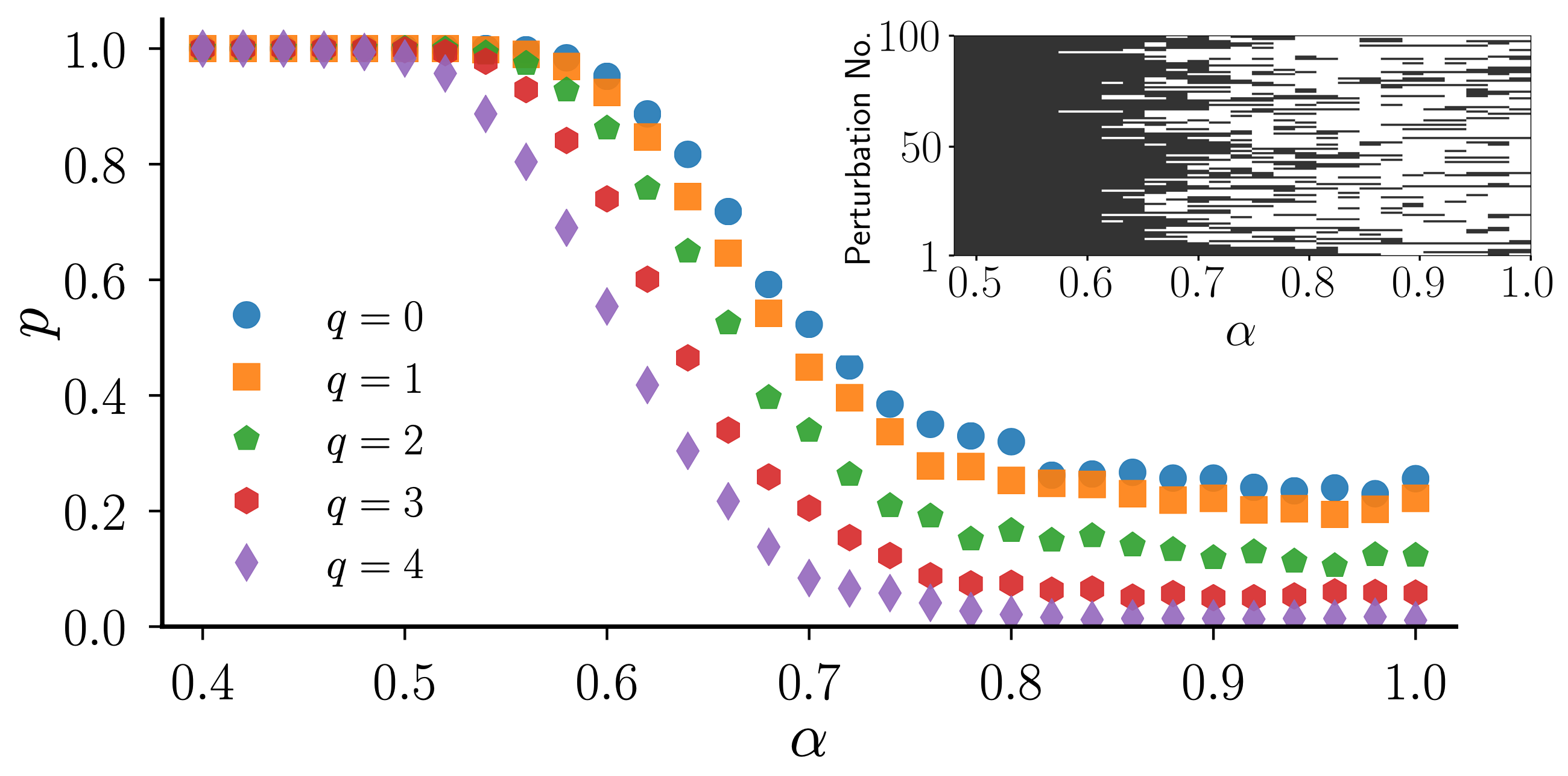}
\vspace{-6mm}
\caption{Probability of returning to a $q$-twisted state after a local perturbation. Each component of the perturbation is selected randomly and uniformly from $[-\alpha\pi,\alpha\pi]$. For all $q$ considered, $p$ stays roughly constant for $\alpha>0.8$. This suggests that the basins are more like octopuses than hypercubes. The inset further confirms this picture, where we show $100$ representative perturbations leaving and reentering the basin of the in-phase state ($q=0$) as $\alpha$ is increased. Here, black indicates inside and white indicates outside of the basin.
}
\label{fig:3}
\end{figure}

To further support the octopus picture of the basins, we scan many different radial lines emanating from twisted states, in an effort to detect rays leaving and reentering the same basin (as one would expect they would, if the basins have tentacles that meander far and wide in state space).
For this purpose, we apply random perturbations to twisted states, whose components are selected independently and uniformly from $[-\alpha\pi,\alpha\pi]$, where $\alpha$ tunes the extent of the perturbation.
In \cref{fig:3}, we vary $\alpha$ from $0.4$ to $1.0$ and estimate the probability $p$ of returning to the original twisted state for different values of $\alpha$.
Interestingly, for all $q$ considered, $p$ stays roughly constant around a nonzero value for $\alpha>0.8$.
This result seems to contradict the dashed curves in \cref{fig:2}, which suggest that almost all rays leave the original basin after a certain distance threshold is crossed.
This apparent contradiction is resolved once one realizes that the rays can reenter the same basin when they are farther away, 
provided that the basins are equipped with tentacles like an octopus instead of being convex like a hypercube.

This intuitive picture of rays repeatedly exiting and reentering the same basin is further supported by the plot shown in the inset of \cref{fig:3}, which depicts the $\alpha$-dependent convergence back to the in-phase state ($q=0$) for $100$ representative rays.
We see that no ray is inside the basin for all $\alpha$.
But for each $\alpha$, there are always rays that are inside the basin.
Moreover, if a ray leaves the basin at a certain value of $\alpha$, it often reenters the same basin at a larger value of $\alpha$.

{\YZ In fact, such repeated reentries 
can be seen as the defining feature of basins with tentacles.
For any given fixed-point attractor, we say its basin is {\it octopus-like} if there is a nonzero probability that a ray emanating from the attractor along a random direction intersects the basin at disjoint intervals.
Note that an octopus-like basin is necessarily concave, but not all concave basins are octopus-like.
Moreover, 
we say an octopus-like basin has {\it long tentacles} if 
the basin cannot be confined within any hypercubes other than the full state space, which is the case for the Kuramoto systems studied here.
}



\begin{figure}[t]
\centering
\includegraphics[width=1\columnwidth]{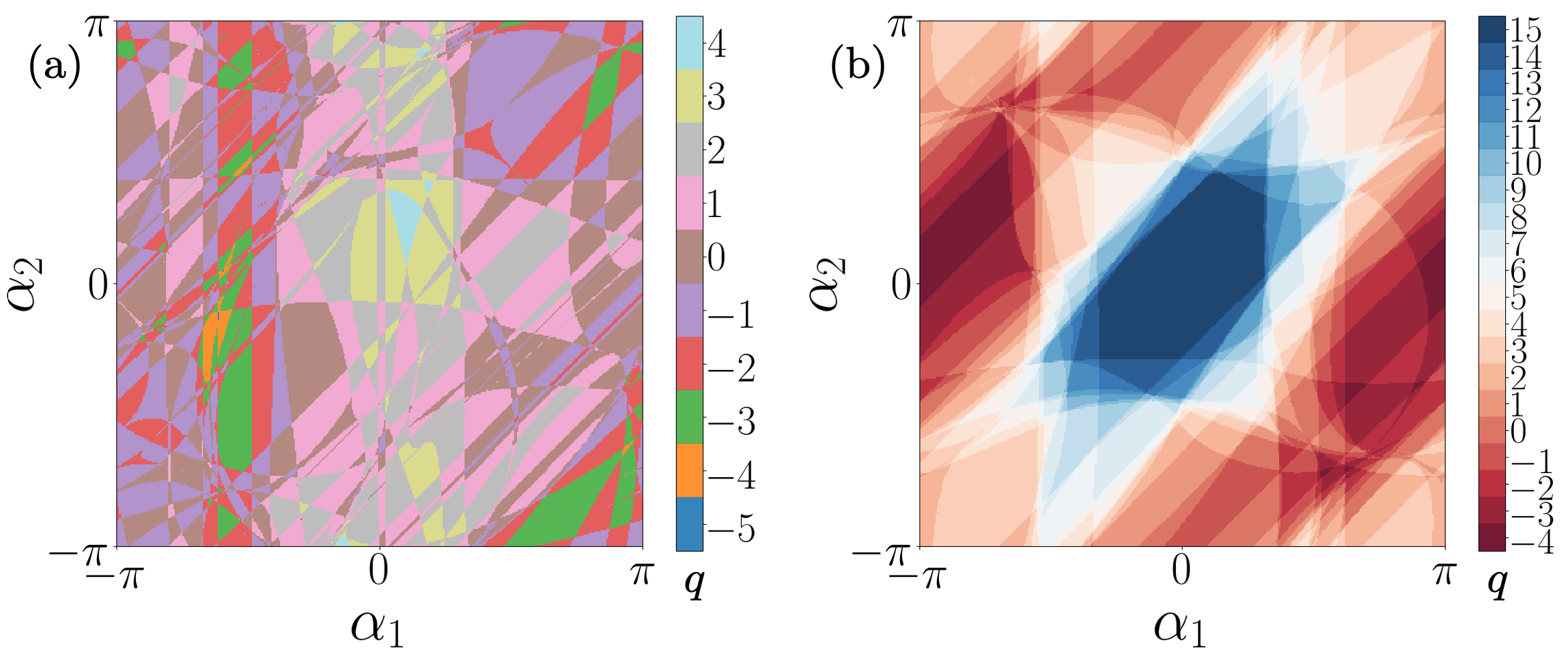}
\vspace{-6mm}
\caption{\YZ Two-dimensional slices of state space reveal the intricacy of basin geometry. (a) Random slice. Basins are color coded by the winding number $q$ of the corresponding attractor. 
The basins appear fragmented. 
(b) Slice centered at the twisted state with $q=15$. 
The color scheme highlights the onion-like structure of the basins; the core (basin for $q=15$) is wrapped inside many layers corresponding to basins with gradually decreasing $q$.
}
\label{fig:4}
\end{figure}

Figure~\ref{fig:4} is a further attempt to visualize the structure of high-dimensional basins, now by examining randomly oriented two-dimensional (2D) slices of state space, either far from a twisted state or close to one. Specifically, we look at slices spanned by $\bm{\theta}_0 + \alpha_1\bm{P}_1 + \alpha_2\bm{P}_2$, $\alpha_i \in (-\pi,\pi]$.
Here, $\bm{P}_1$ and $\bm{P}_2$ are $n$-dimensional binary  orientation vectors in which $\floor*{n/2}$ randomly selected components are $1$ and the rest of the components are $0$.
The results below are not sensitive to the particular realizations of $\bm{P}_1$ and $\bm{P}_2$. However, the choice of the base point $\bm{\theta}_0$ matters a great deal. For example, in \cref{fig:4}(a), we choose $\bm{\theta}_0$ to be a random point in the state space. Despite the fact that each basin is connected (because the dynamics are described by differential equations), the basins look  fragmented in this 2D slice. Perhaps another metaphor than tentacles---a ball of tangled yarn---better captures the essence of the basin structure in this regime, far from any attractor, in which differently colored threads (representing different basins) are interwoven together in an irregular fashion. As one might expect, a random slice of the state space such as this one is dominated by basins corresponding to small values of  $|q|$.

The basin structure near an attractor is strikingly different. In \cref{fig:4}(b), we set $\bm{\theta}_0$ to be the twisted state with $q=15$.
Here, the central basin ($q=15$) is surrounded by competing basins in a structured fashion. As made evident by the color scheme, the basins near an attractor are organized like an onion.
As we peel away the onion layer by layer, the winding number of the basin gradually increases and finally reaches $q=15$ at its core.
(Although we know from above that there must be holes in the onion for the ``tentacles'' of the center basin to snake through.)

Finally, we explain
why octopus-like basins should be prevalent in high-dimensional dynamical systems. 
Consider an $n$-dimensional compact state space with side length $L$ in each direction (after suitable rescaling).
We say a basin is {\it boxy} if it can be confined in a hypercube of side length $\ell<L$.
If all basins of a system are boxy for an $\ell$ that does not depend on $n$, then to fill the entire state space we need at least $(L/\ell)^n$ different basins. So if the number of attractors in a system grows sub-exponentially with $n$, the basins cannot all be boxy. In particular, this is true for the Kuramoto systems we consider here, whose number of attractors grows linearly with $n$. 

Basins can be non-boxy because they are octopus-like, with {\YZ long} tentacles that slither throughout state space and escape any potentially confining hypercube.  
But other scenarios can also occur. Imagine a limiting case where the head of the octopus expands to engulf the tentacles; then the basins  stretch continuously across state space in some or all directions (as they do, for instance, in a system with just one attractor). 
Nevertheless, we predict that basins with tentacles are generic for  high-dimensional dynamical systems with a modest number of attractors, because they provide the least constrained way to fill the state space.
Our prediction is supported by studies on basins in diverse physical systems \cite{monteforte2012dynamic,ashwin2012calculations,martiniani2016structural}, from neuronal circuits to jammed sphere packings. In some cases one can already visually identify tentacles in low-dimensional slices of the basins \cite{menck2014dead,zhiyenbayev2019enhanced}.

By illuminating the structure of octopus-like basins and establishing their prevalence, we hope this work will motivate future studies of basin structure in high-dimensional systems.
{\YZ Some promising directions include the definition of octopus-like basins for chaotic attractors,
understanding the role of saddles in creating basin tentacles,
and generating new insights on reservoir computers \cite{pathak2018model} and adversarial examples in neural networks \cite{goodfellow2014explaining} by characterizing their basin geometries.}


\begin{acknowledgments}
We thank Stefano Martiniani for sharing insights and references on basin structures in jammed sphere packings and Robin Delabays and Ralph Andrzejak for stimulating discussions.
Y.Z. acknowledges support from the Schmidt Science Fellowship and the Santa Fe Institute.
\end{acknowledgments}

\bibliography{ref1,ref2}

\end{document}